\documentclass[a4paper,
               biblatex,     
               ]{jacow}
\usepackage{pdfpages,multirow,ragged2e} %
\usepackage{siunitx}
\makeatletter%
	\ifboolexpr{bool{xetex}}
	 {\renewcommand{\Gin@extensions}{.pdf,%
	                    .png,.jpg,.bmp,.pict,.tif,.psd,.mac,.sga,.tga,.gif,%
	                    .eps,.ps,%
	                    }}{}
\makeatother

\ifboolexpr{bool{xetex} or bool{luatex}} 
 {}                                      
 {\usepackage[utf8]{inputenc}}           

\usepackage{subcaption}
\usepackage[acronym]{glossaries}
\usepackage{tabularx}
\newacronym{knn}{KNN}{K-Nearest Neighbours}
\newacronym{acct}{ACCT}{AC Current Transformer}
\newacronym{epics}{EPICS}{Experimental Physics and Industrial Control System}
\newacronym{fpga}{FPGA}{Field-Programmable Gate Array}
\newacronym{pca}{PCA}{Principal Component Analysis}
\newacronym{pfa}{PFA}{Principal Feature Analysis}
\newacronym{tsne}{t-SNE}{t-Distributed Stochastic Neighbor Embedding}
\newacronym{dbscan}{DBSCAN}{Density-Based Spatial Clustering of Applications with Noise}
\newacronym{optics}{OPTICS}{Ordering Points To Identify the Clustering Structure}
\newacronym{hdbscan}{HDBSCAN}{Hierarchical Density-Based Spatial Clustering of Applications with Noise}
\newacronym{adu}{ADU}{analog-to-digital unit}
\newacronym{pid}{PID}{Proportional-integral-derivative}
\newacronym{plc}{PLC}{Programmable Logic Controller}
\newacronym{rf}{RF}{radiofrequency}
\newacronym{llrf}{LLRF}{Low-Level Radio Frequency}
\newacronym{ioc}{IOC}{Input/Output Controller}

\begin{document}

\title{Development of fault identification pipeline \\for SPIRAL2 LLRF data}

\author{C. Lassalle\textsuperscript{1}, M. Di Giacomo, A. Ghribi, Grand Accélérateur National d'Ions Lourds, Caen, France \\
		P. Bonnay, Université Grenoble-Alpes, CEA IRIG, DSBT, Grenoble, France \\
		F. Bouly, Université Grenoble Alpes, CNRS, Grenoble INP, LPSC-IN2P3, Grenoble, France \\
		\textsuperscript{1}also at Université de Caen Normandie, Caen, France}
	
\maketitle
\begin{abstract}
SPIRAL2 is a state-of-the-art superconducting linear accelerator for heavy ions. The radiofrequency operation of the linac can be disrupted by anomalies that affect its reliability. This work leverages fast, multivariate time series postmortem data from the \gls{llrf} systems to differentiate anomaly groups. However, interpreting these anomalies traditionally relies on expert analysis, with certain behaviours remaining obscure even to experienced observers. By adopting the Time2Feat pipeline, this study explores the interpretability of anomalies through feature selection, paving the way for real-time state observers. Clustering dashboards are presented, allowing the use of multiple clustering algorithms easily configurable and tools to help for visualizing results. A case study on distinguishing electronic quenches and false quench alarms in postmortem data is highlighted. Thereby, a fast and reliable \gls{knn} classifier is proposed.
\end{abstract}

\section{Introduction}
Anomalies in time series are the subject of research in many fields.
These topics can be divided into subgroups:
\begin{itemize}
	\item detection: identify behaviours that deviate from nominal operating modes.
	\item classification: recognize the types of anomalies.
	\item localization\cite{shimillas2025transformerbasedmultivariatetimeseries}: distinguish the signals causing these malfunctions.
\end{itemize}

Particle accelerators are no exception to these issues, as they are complex systems that must be kept running for users. During operation, some events such as quenches, multipacting, or microphonics, may occur, leading to beam loss and reduced accelerator availability. Exploring and understanding these events is crucial to enhance the reliability of the accelerator. For CEBAF, work focusing on particle orbit has been carried out \cite{chen2024anomalydetectionparticleorbit}. Other work has been directed toward the classification of faults \cite{PhysRevAccelBeams.23.114601}.
Similarly, research on this topic has been undertaken for CAFE2 \cite{yang_classification_2025}.
For EuXFEL, the focus was on identifying quenches \cite{boukela2024twostagemachinelearningaidedapproach}.\\

\begin{figure}
\centering
\begin{subfigure}{0.4\linewidth}
    \includegraphics[width=\linewidth]{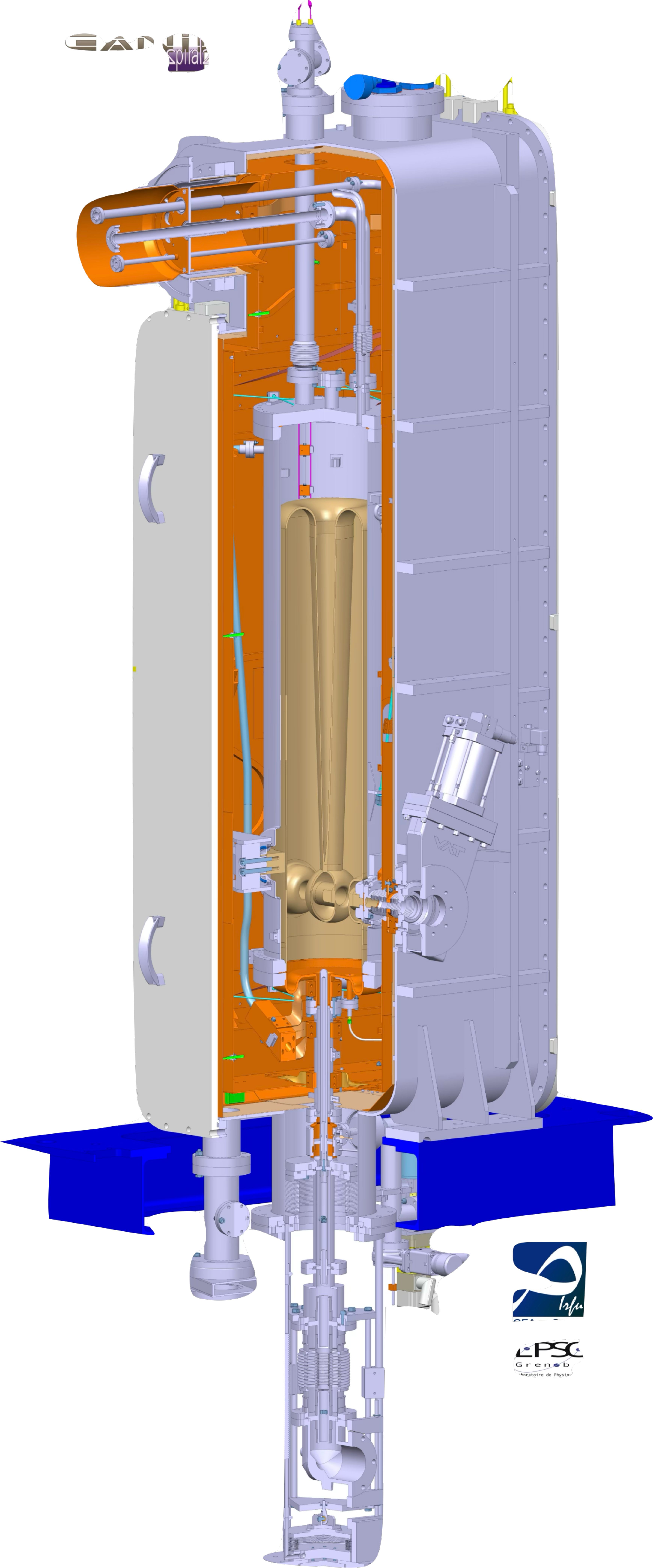}
    \caption{Type A}
    \label{fig:cryomoduleA}
\end{subfigure}
\hfill
\begin{subfigure}{0.4\linewidth}
    \includegraphics[width=\linewidth]{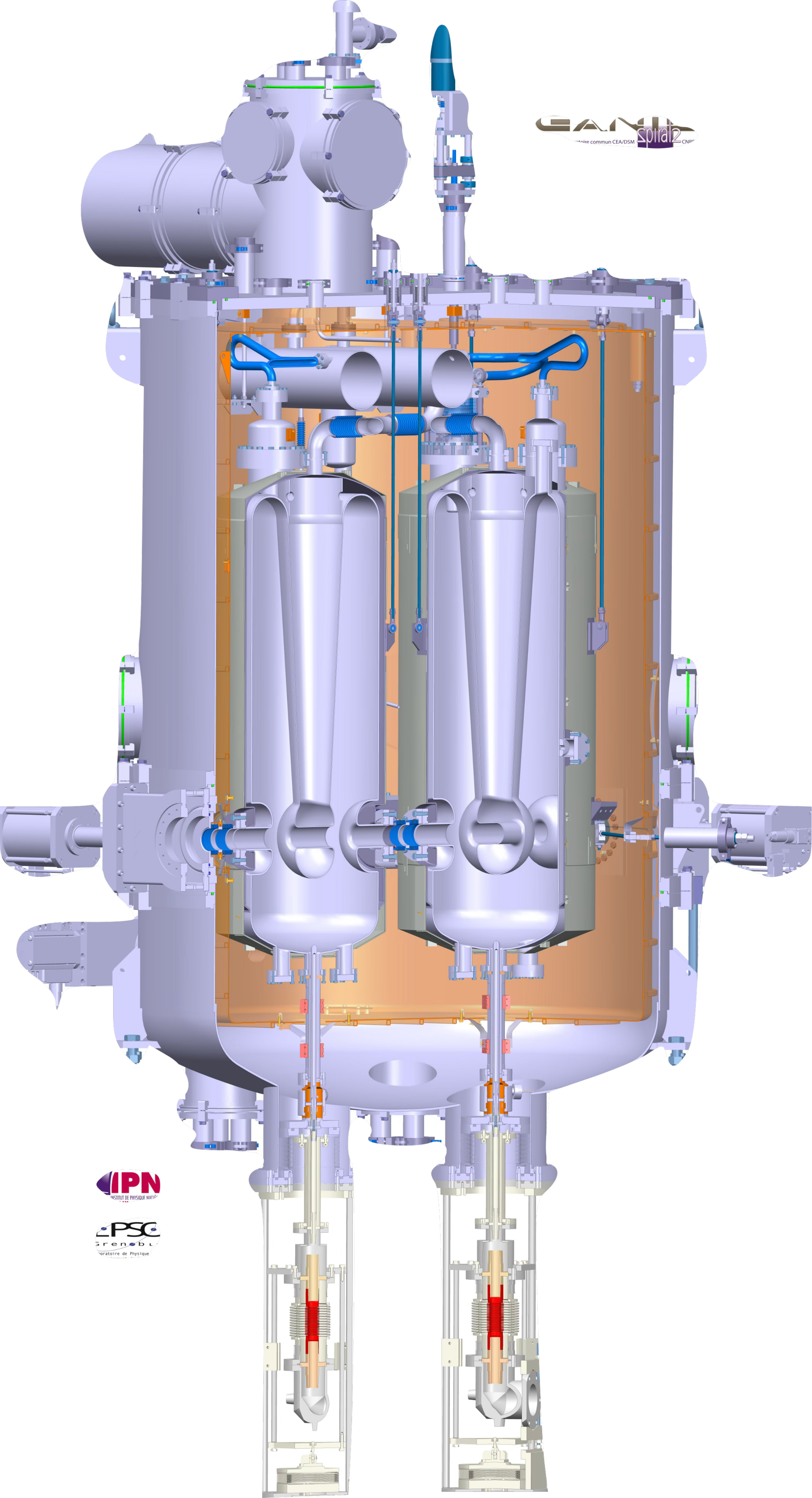}
    \caption{Type B}
    \label{fig:cryomoduleB}
\end{subfigure}
        
\caption{SPIRAL2 cryomodules.}
\label{fig:cavities}
\end{figure}

This paper presents the initial efforts to analyse and apply machine learning methods to SPIRAL2 \gls{llrf} data, aiming to identify and classify faults.
Located at GANIL (Caen, France), this linear accelerator is dedicated to the production of rare and exotic ion beams for nuclear physics research \cite{orduz2024spiral2}. A cryogenic plant supplies the liquid helium at \qty{4.2}{\kelvin} needed to cool its 26 superconducting niobium cavities (Fig.~\ref{fig:cavities}). These are powered by solid-state amplifiers with a maximum power of \SI{10}{\kilo\watt} for the low beta ones and \SI{20}{\kilo\watt} for all the high beta ones.\\

In the first section, we introduce the structure of the data, then we present the tools used for its exploration. Next, the main methodology is explicited and a case study on e-quenches \cite{powers:srf93-srf93k01} events is included. Finally, we suggest some perspectives for this work.
\section{LLRF \& acquisition system}
\begin{figure}
	\centering
	\includegraphics[width=\columnwidth]{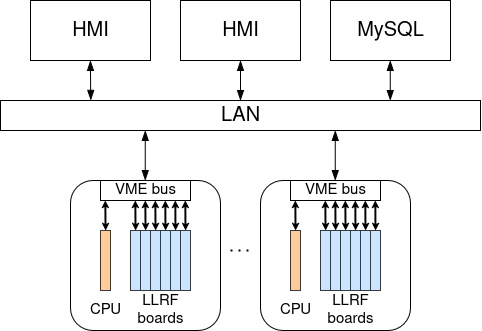}
	\caption{Acquisition system overview.}
	\label{fig:acquisitionsystem}
\end{figure}

Each cavity is associated with a digital LLRF board, based on a \gls{fpga}. These boards are connected to an \gls{epics} \gls{ioc} for monitoring and control purposes through the local ethernet network (Fig.~\ref{fig:acquisitionsystem}) and are equipped with circular memory. When an alarm is triggered according to predefined criteria, a set of signals is stored in binary or ASCII files. The binary format is preferable because it requires less storage space. Data can also be recorded manually by an operator. The acquisition can be configured in terms of number of samples, sampling frequency, and pre-trigger duration. The sampling of the values can go down to \qty{110}{\nano\second} but at the expense of the duration of the event. Most of the files contain values recorded every \qty{11}{\micro\second} and are centred around the moment of manual triggering or the alarm. These faulty events are used in this work.\\

Historically, postmortem data were processed with a proprietary code and only graphical interpretation from time-domain graphs was possible. In order to progress towards more FAIR (Findable Accessible Interoperable Reproducible) compatible datasets, a different open-source and python compatible preprocessing has been implemented, including metadata management and HDF5 exports.

\section{Data structure}
Each file follows the same structure: a header and multivariate time series, including signals, faults and states.

\begin{description}
    \item[The header] The header contains metadata. In particular, it includes the date of the event, the affected cavity, the alarms triggered, the set points, and the calibration values.
    \item[Multivariate time series]
The signals recorded in \gls{adu} are as follows:
\begin{itemize}
 \item Cavity field (in-phase and quadrature components)
 \item Frequency reference signal (in-phase and quadrature components)
 \item Forward power (in-phase and quadrature components)
 \item Reflected power (amplitude)
 \item Input signal of solid-state amplifier (amplitude)
 \item I/Q modulator command (in-phase and quadrature components)
 \item Vacuum pressure
 \item Electron pick-up coupler probe current 
\end{itemize}

When reading files, signals are calibrated and amplitude/phase conversions are performed. The cavity phase shift is also computed.
\item[Faults] The alarms triggered and their relative conditions are presented in the Table \ref{tab:llrfalarms}.

\begin{table}[]
    \centering
        \caption{LLRF Alarms}
    \begin{tabularx}{\linewidth}{lX} 
 \toprule
	\textbf{Faults} & \textbf{Triggering conditions} \\
    \midrule
 Electron pick-up & Pickup current less than threshold\\
Fast interlock & Arc detected by photodiode in circulator \\
No RF permission & No PLC permission\\
Vacuum threshold & Power coupler pressure greater than threshold \\
Cavity breakdown or quench & Amplitude of cavity field drops by $50\%$ in less than \qty{2}{\micro\second}\\
RF protection threshold & Amplitude of reflected power or cavity field more than threshold\\
RF signal out of tolerance & Cavity field undergoes variations greater than \qty{10}{\percent} in amplitude or \qty{1}{\degree} in phase (depending on the set points).\\
\bottomrule
\end{tabularx}
    \label{tab:llrfalarms}
\end{table}

\item[States]The states are additional information. They are not used in this work.
\end{description}
\section{Data exploration}
\begin{figure}[hbt]
	\centering
	\includegraphics[width=\columnwidth]{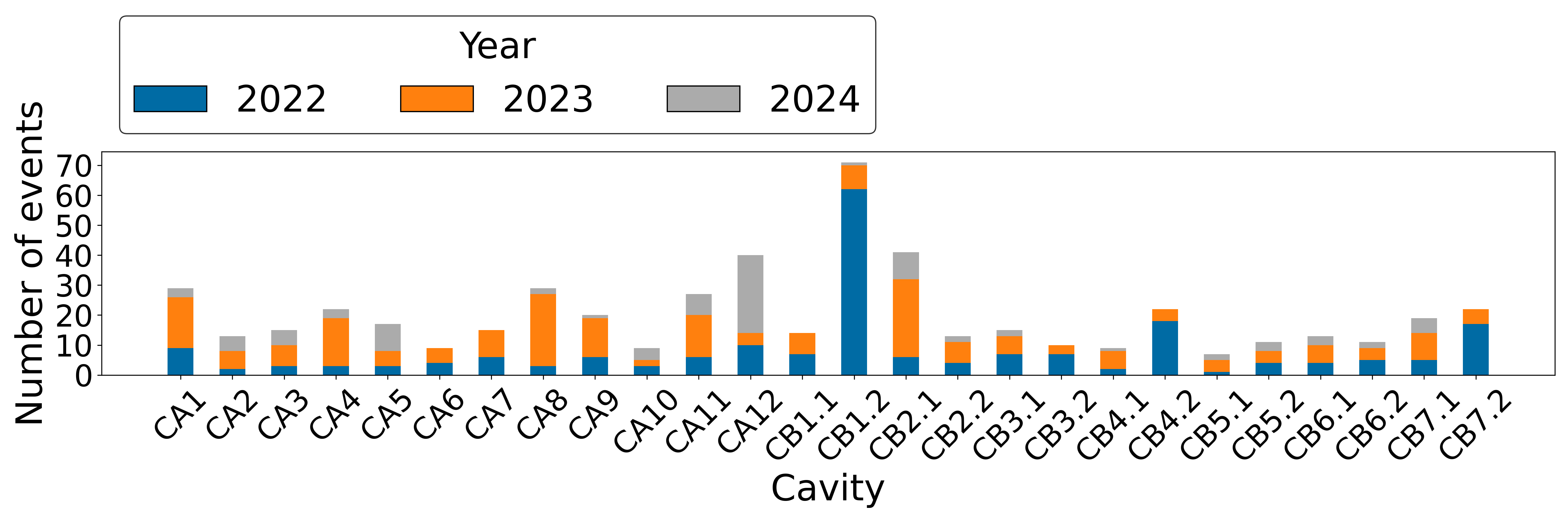}
	\caption{Summary of events by year.}
	\label{fig:events_summary}
\end{figure}

\begin{figure}[hbt]
	\centering
	\includegraphics[width=\columnwidth]{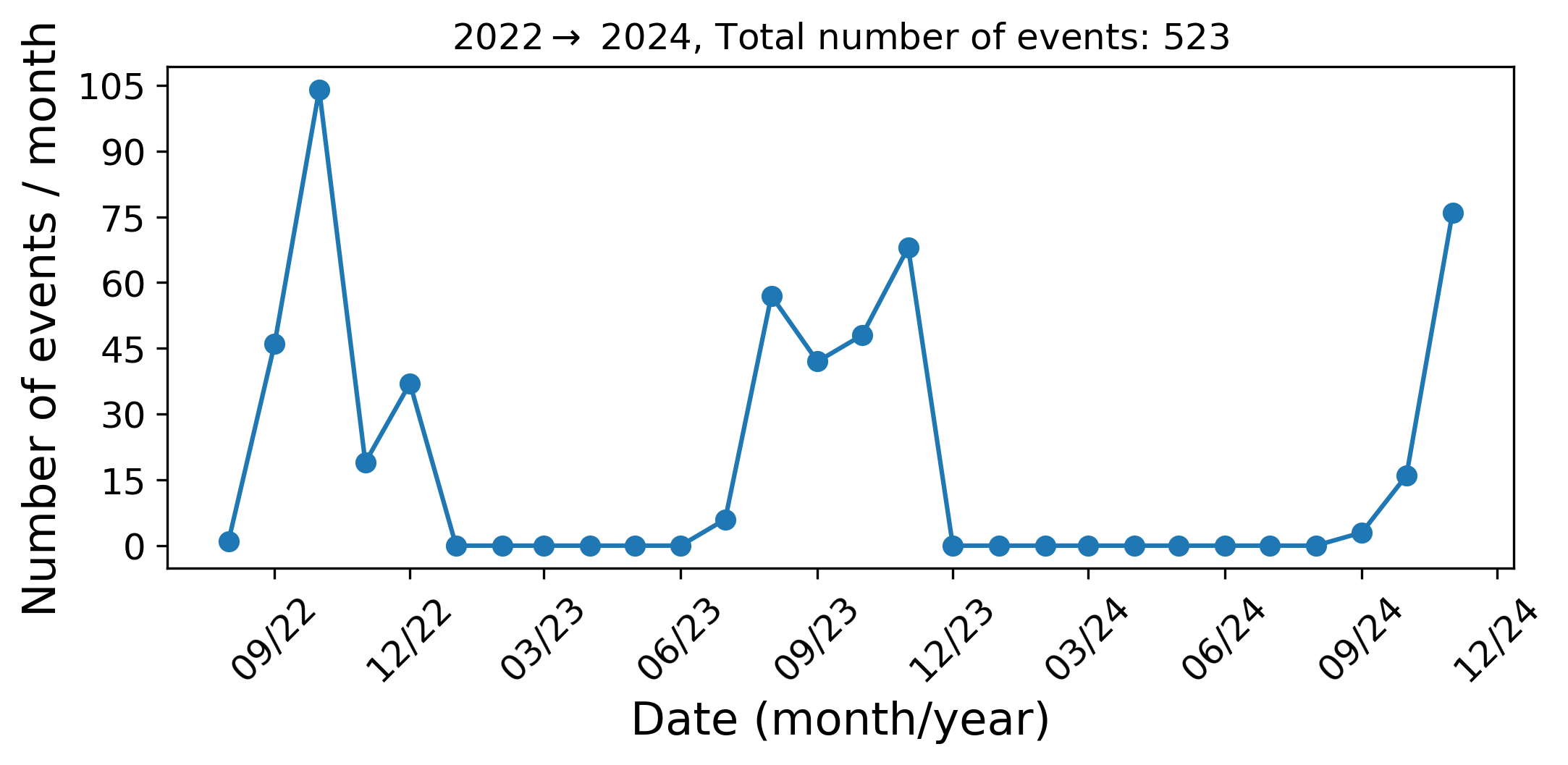}
	\caption{Evolution of alarms over time.}
	\label{fig:events_evolution}
\end{figure}

A first exploration of the data is performed. For that, a dashboard was developed using the Plotly Dash library. The events can be summarized according to various criteria. As illustrated by the Fig.~\ref{fig:events_summary}, the occurrences of faulty events are displayed, by year and for each cavity, with bar charts. This tool is useful for getting an initial overview of the data and seeing how alarms evolve over time, as presented by the Fig.~\ref{fig:events_evolution}.

\section{Clustering of \\multivariate time series}
The signals recorded are processed according to the Time2Feat \cite{time2feat} pipeline. Dashboards tools have been prepared to facilitate configuration.

\subsection{Event Selection}
Events can be selected based on date, beam presence, PID parameters, or the average field present in the cavity before the alarm was triggered. Beam presence is determined with \gls{acct} data stored in a MySQL database (Fig.~\ref{fig:acquisitionsystem}).
\subsection{Preprocessing}
Signals may be truncated to retain only a reduced portion centered around the triggering of the alarm. Smoothing using a moving average can also be applied.
\subsection{Feature Extraction}
From each signal, features such as statistics, variations, or spectral analysis measurements are calculated. The tsfresh library \cite{CHRIST201872}, which combines numerous Python libraries, is used for this step.
\subsection{Feature Selection}
A selection of features is made. First, features with zero variance are removed. Then, a \gls{pfa} retains only those that contribute most to the variance of the data. To do this, a cumulative explained variance threshold is used. 
\subsection{Clustering}
Various clustering algorithms can be used such as k-means, DBSCAN, HDBSCAN or OPTICS. When an algorithm is selected, its hyperparameters can be entered.
\subsection{Visualization of the Results}	
\begin{figure}[hbt]
	\centering
	\includegraphics[width=\columnwidth]{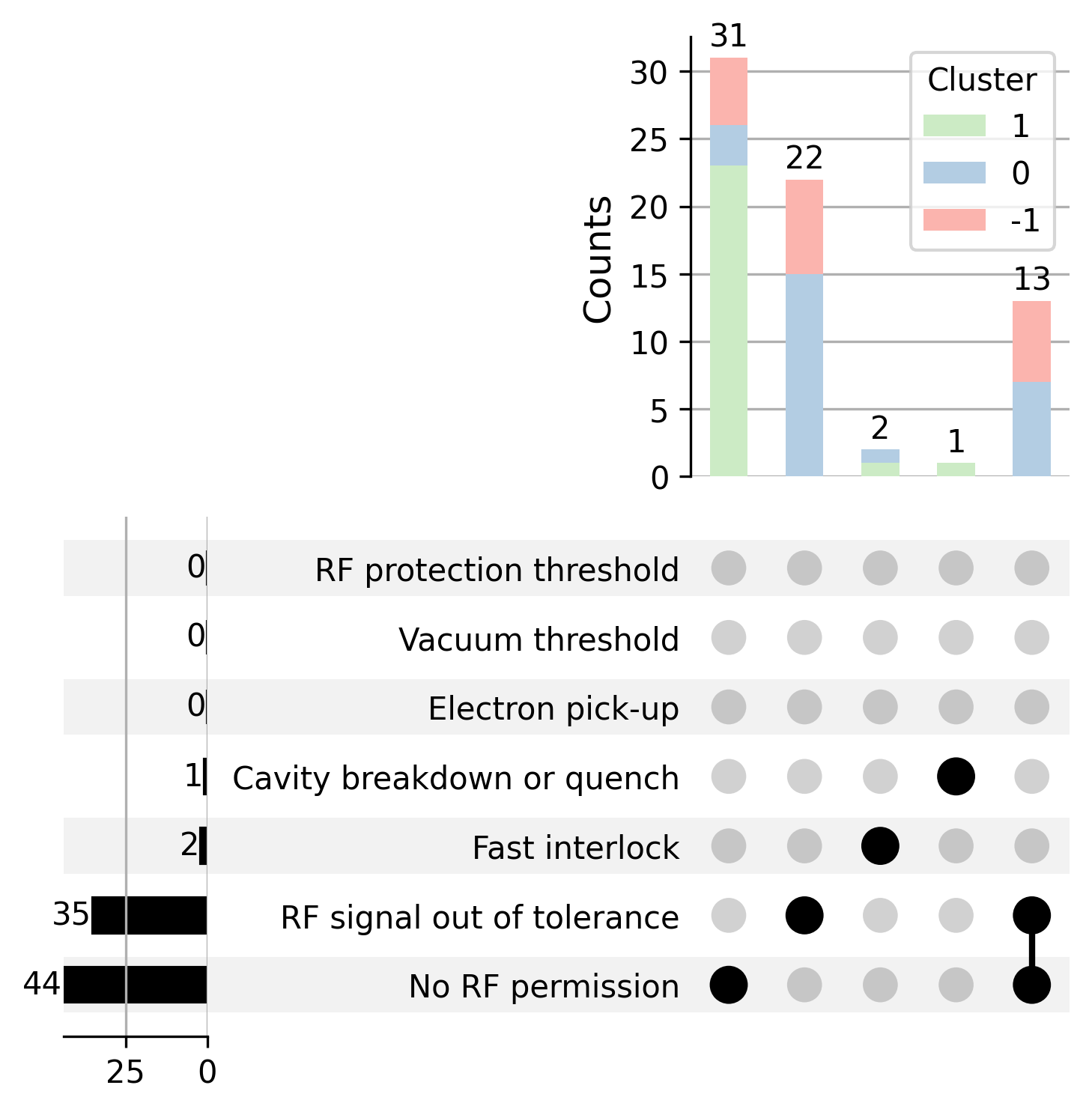}
	\caption{Example of clustering results with upset plot.}
	\label{fig:upsetplot}
\end{figure}

Another dashboard is dedicated to viewing the results of clustering. A representation called upset plot, presented in the Fig.~\ref{fig:upsetplot}, compares the clusters with the \gls{llrf} labels. For each cluster, the list of events is displayed. The corresponding signals can be plotted by selecting files using checkboxes. Fourier transform of chosen signal is available.
\newline

This methodology has not yet demonstrated perfect separation of events, but the tools developed are nevertheless useful and can be improved. Scaling options and a pre-selection of features  could be added.
\section{Case study: \\quench-labeled events}
\begin{figure}[hbt]
\centering
\includegraphics[width=\columnwidth]{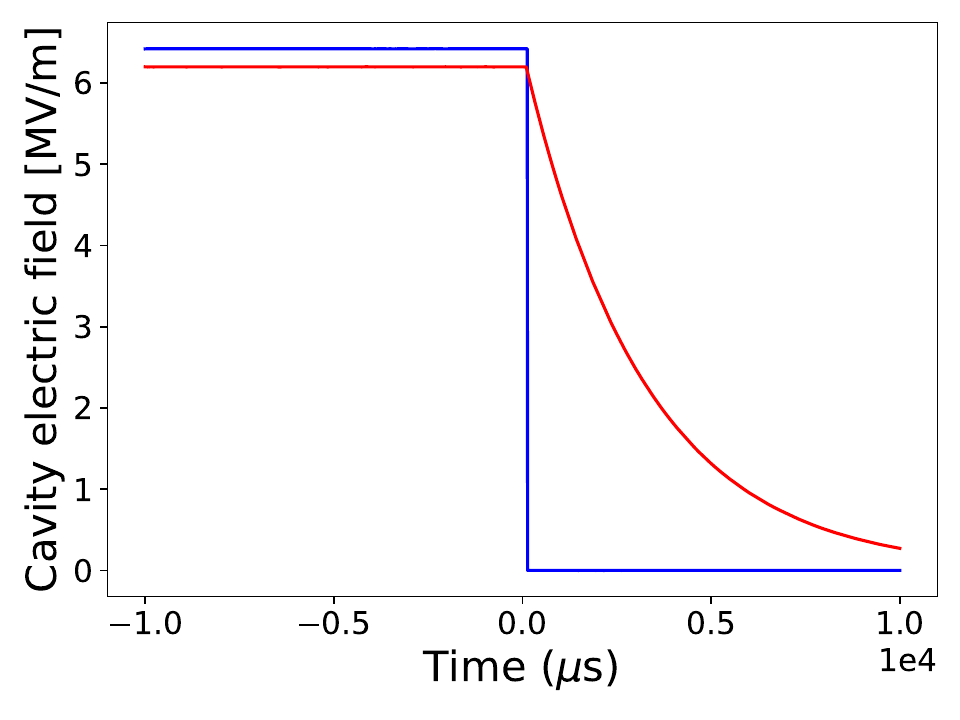}
	\caption{Profiles of events.}
	\label{fig:TS_ucavprofiles}
\end{figure}

The exploration of the data shows that the faults of type ``Cavity quench or breakdown" can be divided into two distinct groups (Fig.~\ref{fig:TS_ucavprofiles}). The first group corresponds to a sudden drop in the electric field in the cavity. This profile is typical of an electronic quench, as observed at CEBAF or CAFe2. The second group corresponds to an exponential decay of the field. However, these events do not correspond to quenches as the time constant is too high. These events are therefore probably false alarms. The reason for their triggering is currently undetermined. It is important to note that other events not labeled as ``Cavity quench or breakdown" by the \gls{llrf} system, actually are. The objective of this case study is to differentiate these two types of events. These events are manually labeled: 24 according to the ``e-quench" category and 11 for the ``false alarm" category. The events are randomly split into training and test sets, with a 70/30 ratio. Using tsfresh, features are extracted from the time series. Only the amplitude of the field in the cavity was considered. The calculated features are then scaled through z-score normalization.
\subsection{Labels Balance}
\begin{figure}[hbt]
	\centering
	\includegraphics[width=\columnwidth]{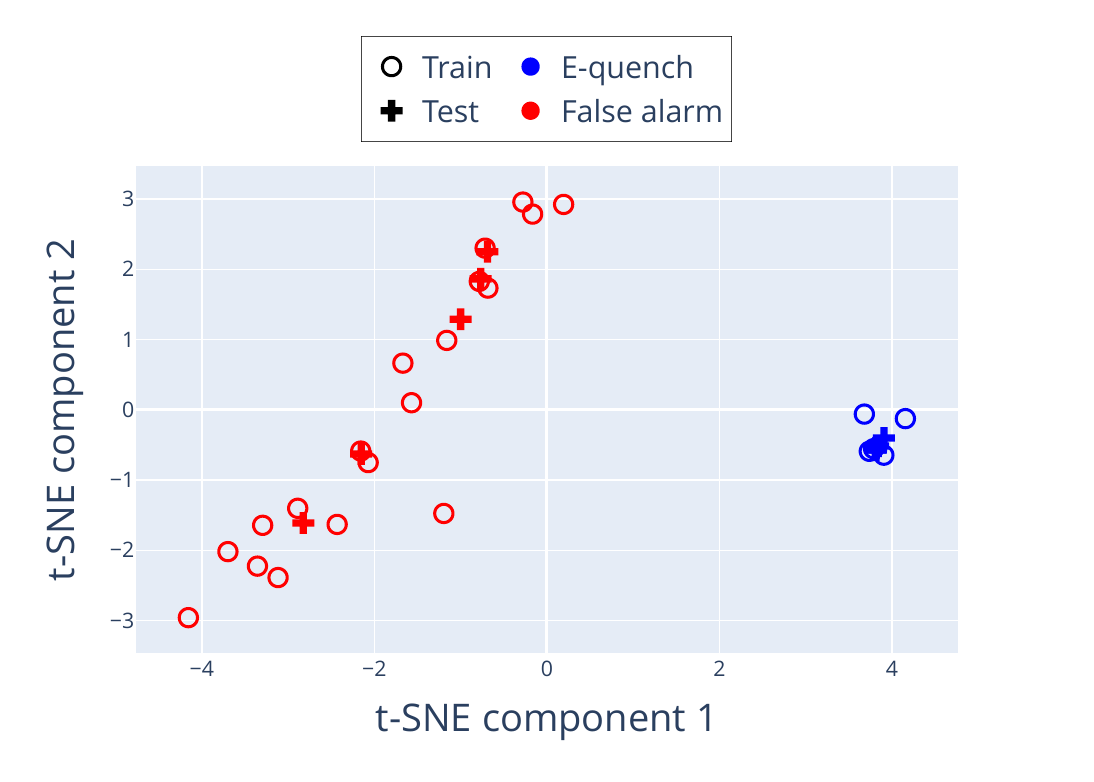}
	\caption{Quench events classification.}
	\label{fig:example_figure}
\end{figure}

\begin{figure}[hbt]
	\centering
	\includegraphics[width=\columnwidth]{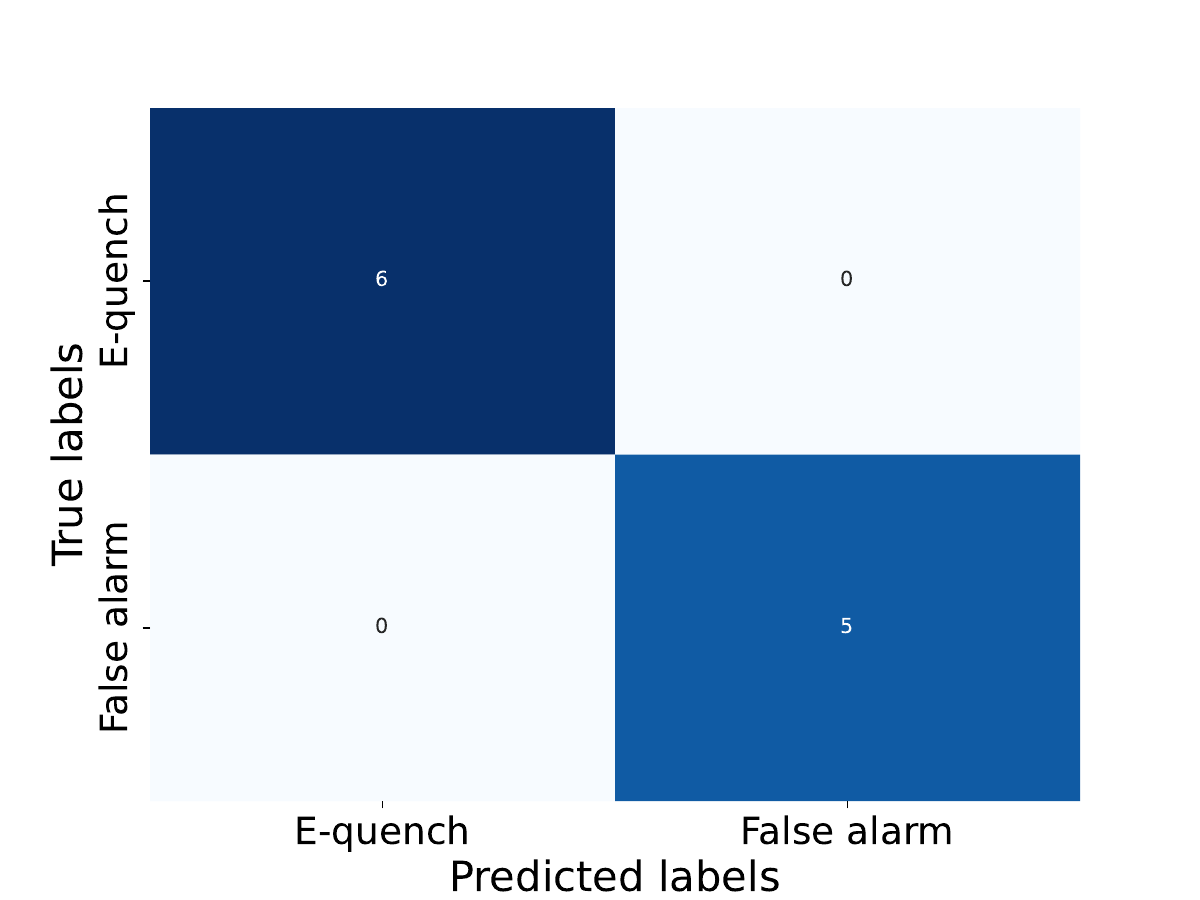}
	\caption{Confusion matrix.}
	\label{fig:confusionmatrix}
\end{figure}

The training set is unbalanced, with 20.83\% of e-quenches and 79.17\% of false quenches events. The test set is also unbalanced, with 54.55\% of ``e-quenches" and 45.45\% of false quenches events. The imbalance is reversed between the two sets. The dimensionality reduction of the training data is performed by t-SNE on computed features. To be able to apply the transformation to the test data, the openTSNE library is used. This library allows to save the t-SNE transformation of the training data and apply it to the test data. The clustering algorithm DBSCAN is then used. Two clusters are formed. The performance of the clustering is evaluated by comparing the predicted labels with the true labels. This verification allows to associate cluster 0 with e-quenches and cluster 1 with false quenches. A \gls{knn} classifier is trained on the training data, then evaluated on the test data (Fig.~\ref{fig:example_figure}). The confusion matrix is shown by the Fig.~\ref{fig:confusionmatrix}. The classifier correctly predicts all test events, even with unbalanced sets. Despite the small number of labeled events, these results are promising and show that the differentiation of these data is possible. Thus, this approach could be used to filter these false alarms in order to save storage memory.
\section{Perspectives}
In addition to improving the ability to differentiate between recorded events, it would be very beneficial to give serious consideration to the interpretability of the results. This would provide assistance to RF experts.
Alternative approaches could be applied, such as the use of autoencoders or transformer neural networks to distinguish between nominal behaviours and faulty ones. Supervised methods, applied to all available data, could be used but would require time-consuming labeling. These algorithms could be implemented in embedded systems, such as \glspl{fpga}, to act in real time.
\section{Conclusion}
Although this first work has not yet led to an effective classification of LLRF data, it paves the way for future research and helps define objectives. Some application cases show encouraging prospects for the use of machine learning on SPIRAL2 \gls{llrf} data.
\section{ACKNOWLEDGEMENTS}
This work has been funded by “Region Normandie” as
well as the city of Caen, CNRS and CEA.
\printbibliography
\end{document}